# Wide-field Decodable Orthogonal Fingerprints of Single Nanoparticles Unlock Multiplexed Digital Assays

Jiayan Liao[1], Jiajia Zhou[1,*], Yiliao Song[2], Baolei Liu[1], Yinghui Chen[1], Fan Wang[1], Chaohao Chen[1], Jun Lin[3], Xueyuan Chen[4], Jie Lu[2,*], Dayong Jin[1,5,*]

[1]Institute for Biomedical Materials and Devices (IBMD), Faculty of Science, University of Technology Sydney, NSW 2007, Australia

[2]Centre for Artificial Intelligence, Faculty of Engineering and IT, University of Technology Sydney, NSW 2007, Australia

[3]State Key Laboratory of Rare Earth Resource Utilization, Changchun Institute of Applied Chemistry, Chinese Academy of Sciences, Changchun, 130022, PR China

[4]CAS Key Laboratory of Design and Assembly of Functional Nanostructures and Fujian Key Laboratory of Nanomaterials, Fujian Institute of Research on the Structure of Matter, Chinese Academy of Sciences, Fuzhou, Fujian 350002, P. R. China

[5]UTS-SUStech Joint Research Centre for Biomedical Materials & Devices, Department of Biomedical Engineering, Southern University of Science and Technology, Shenzhen, China

*e-mail: jiajia.zhou@uts.edu.au; jie.lu@uts.edu.au; dayong.jin@uts.edu.au

**The control in optical uniformity of single nanoparticles and tuning their diversity in orthogonal dimensions, dot to dot, holds the key to unlock nanoscience and applications. Here we report that the time-domain emissive profile from single upconversion nanoparticle, including the rising, decay and peak moment of the excited state population ($\tau^2$ profile), can be arbitrarily tuned by upconversion schemes, including interfacial energy migration, concentration dependency, energy transfer, and isolation of surface quenchers. This allows us to significantly increase the coding capacity at the nanoscale. We exemplify that at least three orthogonal dimensions, including the excitation wavelength, emission colour and $\tau^2$ profiles, can be built into the nanoscale derivative $\tau^2$-Dots. We further implement both time-resolved wide-field imaging and deep-learning techniques to decode these fingerprints, showing high accuracies at high throughput. This work reveals a vast library of individually pre-selectable nano-tags, and our ability for controlling the uniformity and diversity "at the bottom". These high-dimensional optical fingerprints provide a new horizon for applications spanning from sub-diffraction-limit data storage, security inks, to high-throughput single-molecule digital assays and super-resolution imaging.**

"There's Plenty of Room at the Bottom"[1], and it is the ultimate goal of nanotechnology to manipulate structures with unprecedented accuracy and to tune their functions to precisely match the parameters required at single nanoparticle level[2]. Optical multiplexing with increased capacity will advance the ongoing development of next-generation enabling technologies, spanning from high-capacity data storage[3], anti-counterfeiting[4–7], large-volume information communication[8,9], to high-throughput screening of multiple single molecular analytes in a single test[10,11], and super-resolution imaging of multiple cellular compartments[12–14]. Super-capacity optical multiplexing challenges our abilities in creating multiplexed codes in orthogonal dimensions, e.g. intensity, colour, polarization, and decay time, assigning them to the microscopic and nanoscale carriers, and decoding them in high throughput fashion with sufficient accuracy in the orthogonal optical dimensions[15,16]. Though the desirable size of material that carries the optical barcodes can be pushed from microscopic to the nanoscopic range, it sacrifices the overall amount of emissive photons (brightness) and therefore limits the number of detectable codes, e.g. typically three to four colour channels or brightness levels[17]. The amount of signal from a nanoscale object can drop exponentially and the size of them is often below the optical diffraction limit, which prevents the conventional filter optics and detection process from decoding them with sufficient spectral-spatial resolutions.

The unmet expectation poses grand challenges for the material sciences to pursue the fabrication strategies and the precise control in producing uniform nanoscopic carriers, and further challenges the photonics community to maximize the number of emissive photons and to explore the diversity of optical information that can be produced in multiple orthogonal dimensions, such as emission colours (spectrum)[18,19], lifetime[20–24], polarization[25–28], and angular momentum[29,30]. The goal is to arbitrarily assemble these orthogonal codes within a shared space at the nanoscale, such as the building block of single nanoparticles[2].

Lanthanides doped upconversion nanoparticles (UCNPs) absorb the low-energy near-infrared (NIR) photons to emit high-energy emissions in visible and UV regions[31]. Single UCNPs are uniform, photo-stable for hours, and allows single nanoparticle tracking experiments in live cells[32–34]. Recently, the core-shell-shell design of each single UCNP has been reported can emit ~ 200 photons per second under a low irradiance of 8 W/cm$^2$ [35], and intensity uniform UCNPs have enabled the single-molecule (digital) immuno assay[36,37]. The colour-based multiplexing of UCNPs can be realized by tuning the dopants[38–40], core-shell structure[41,42], or excitation pulse durations[43], but all the colour-based approaches are intrinsically limited by the cross-talks in the spectrum domain. Recently, we tuned the exceptional long lifetimes of UCNPs' in a microsecond-to-millisecond range for multiplexing applications[21]. Major advances have since been made in the ensemble lifetime measurements of microsphere arrays[44], time-domain contrast agents for deep-tissue tumour imaging[45] and high-security-level anticounterfeiting applications[46]. Though lifetime multiplexing with single nanoparticle sensitivity was possible, the relatively low brightness and the point scanning modality of confocal microscopy have limited the readout throughput.

In this work, we first demonstrate that the morphology of both the typical structure of active core @ inert shell UCNPs and the rather sophisticated design of active core @ energy migration shell @ sensitization shell @ inert shell UCNPs can be highly controlled. We find that once the single UCNP is sufficiently bright under wide-field excitation, it displays its' characteristic optical signatures in the time domain. Remarkably, not only the decay time of each batch of UCNPs is tunable, but also the rising time and peak moment of the excited state population from a single nanoparticle can be further manipulated by multi-interfacial energy transfer process and orthogonal excitation wavelengths. We emphasize that UCNPs from each batch of controlled synthesis can carry a unique set of optical signatures in multiple dimensions. Any new dimension including the rising time and peak moment, as the new differentiators to be detected, can build new levels of confidence in distinguishing one batch from another, and these high-dimensional optical signatures can be pre-selected to build a vast library of single-particle tags. This finding provides a new opportunity for super-capacity multiplexing that integrates multiple orthogonal dimensions of signals from single nanoparticles.

## Results

### Control of $\tau^2$ profile

Three typical series of UCNPs have been investigated in this work (Table S1), displaying three orthogonal dimensions (excitation wavelength, emission wavelength, and lifetime) of optical fingerprints. The Yb-Tm series (Figure 1a-1c) can be excited at 976 nm, the Nd-Yb-Er series (Figure 1d-1g) and Nd-Yb-Tm (Fig. S1) allow both 976 nm and 808 nm laser excitations. The TEM images in Figure 1a and Fig. S2 shows the uniform spherical β-NaYF$_4$: Yb$^{3+}$, Tm$^{3+}$ core @ inert shell nanoparticles (coefficients of variation (CV) < 5%). Upon excitation of 976 nm, the nanoparticles emit in blue, red and NIR spectral bands, which are assigned to the diverse transitions of Tm$^{3+}$ (Figure 1b). The whole band emissions from the excited states ($^1G_4$, $^1D_2$, $^3H_4$) exhibit overall lifetime profile including both rising and decay components on a microsecond time scale (Figure 1c). This profile renders each different batch of nanoparticles a unique optical fingerprint, featured by a rather sophisticated multi-component lifetime behaviour, namely $\tau^2$-Dot profile.

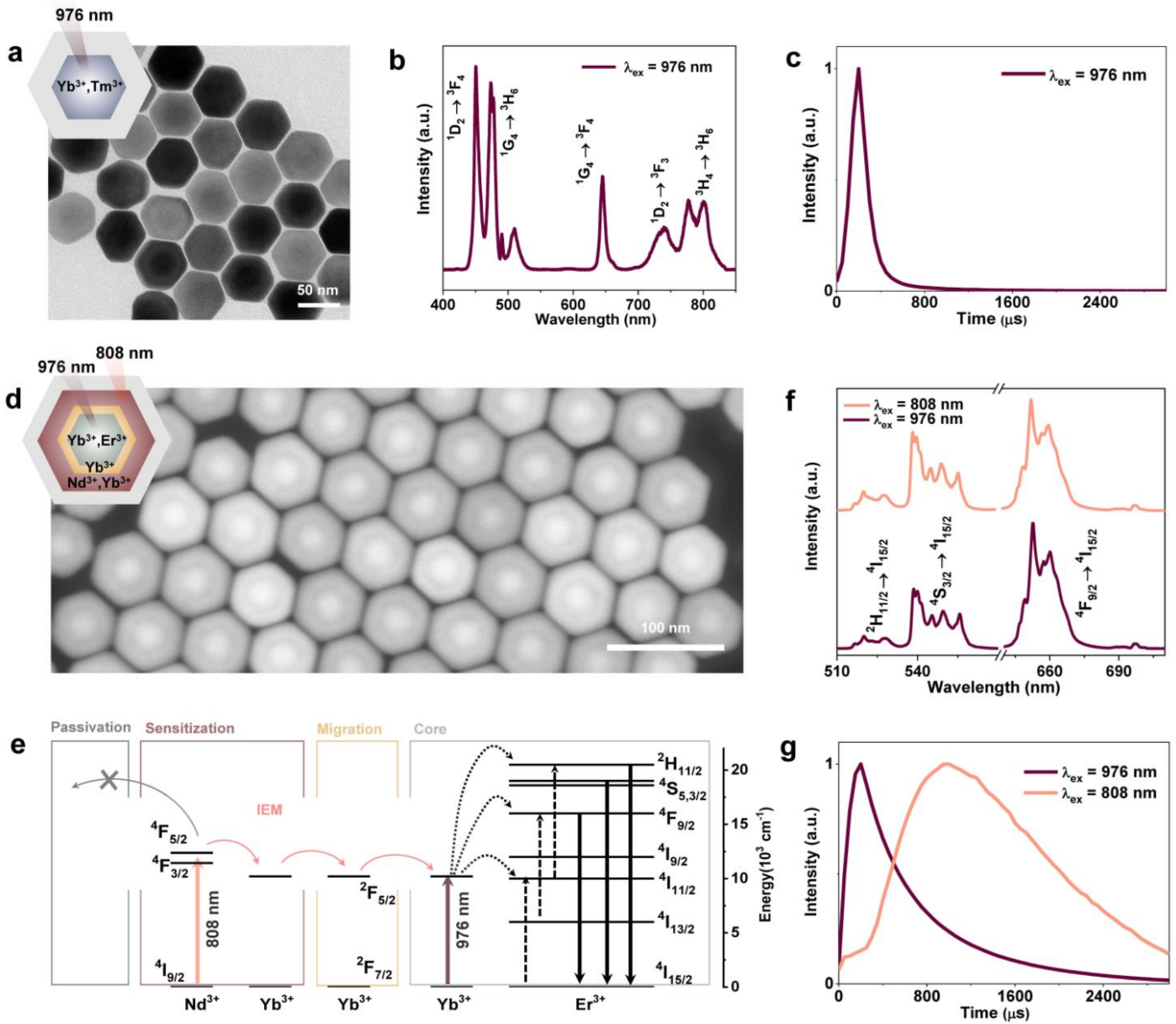

**Fig. 1 | Creation of monodisperse UCNPs with optical information in orthogonal dimensions. (a)** TEM image of a kind of typical morphology uniform core-shell nanoparticles β-NaYF$_4$: Yb$^{3+}$, Tm$^{3+}$. **(b and c)** Ensemble upconversion emission spectrum (b) and lifetime profile (c) from 400 to 750 nm of the Yb$^{3+}$, Tm$^{3+}$ doped UCNPs under 976 nm excitation. **(d)** HTEM observation showing the core-multi-shell structure of the nanoparticles doped with Nd$^{3+}$, Yb$^{3+}$, Er$^{3+}$ in different layers. **(e)** Energy level diagram of core-multi-shell nanoparticles showing the cascade photon energy sensitization, transfer and conversion process: Nd$^{3+}$ sensitization at 808 nm, Yb$^{3+}$-mediated interfacial energy migration (IEM) at 976 nm, and upconversion of near-infrared photons into higher-energy visible emissions in a typical Yb$^{3+}$-Er$^{3+}$ system. **(f and g)** Ensemble upconversion emission spectra (f) and lifetime curves (g) from 400 to 750 nm of Nd$^{3+}$, Yb$^{3+}$, Er$^{3+}$ doped UCNPs under 808 nm and 976 nm excitations. The pulse width of 200 μs was used for all the lifetime measurements.

**The role of interfacial energy migration in lifetime engineering**

The τ$^2$-Dots concept can be further extended by making a series of core-multi-shell β-NaYF$_4$: Nd$^{3+}$, Yb$^{3+}$, Tm$^{3+}$ UCNPs (Fig. S3) and β-NaYF$_4$: Nd$^{3+}$, Yb$^{3+}$, Er$^{3+}$ UCNPs (Figure 1d and Fig. S4) with a great morphology uniformity (CV < 5%). The sophisticated design of core-multi-shell UCNPs permits an arbitrary control in the energy transfer process within a single nanoparticle[47], as illustrated in Figure 1e: the shell co-doped with $Nd^{3+}$ and $Yb^{3+}$ ions sensitizes 808 nm excitation, the energy migration shell containing a small percentage of $Yb^{3+}$ ions is responsible for passing on the absorbed energy to the conventional $Yb^{3+}$, $Er^{3+}$ co-doped core that emits up-converted emissions at green and red bands (see Figure 1f), and an inert shell is employed to prevent the energy migration to the surface quenchers[48,49], as well as improving the optical uniformity of single nanoparticles. The multiple shells can significantly slow down the interfacial energy migration (IEM) process from primary sensitizer $Nd^{3+}$ to the secondary sensitizer $Yb^{3+}$ under the excitation of 808 nm. IEM plays an important role in the slow accumulation of the excited state populations, displayed as a time-delayed up-rising curve of upconversion emissions. To verify this IEM effect, we selectively excite the $Yb^{3+}$ and $Nd^{3+}$ ions using 976 nm and 808 nm lasers, respectively, and observe the same emission spectra (Figure 1f), but much differences in the $\tau^2$ profiles (Figure 1g). The rising time for the $Er^{3+}$ excited state populations to reach its plateau can be prolonged from 200 μs to 950 μs when IEM process is involved.

**Orthogonal optical fingerprint encoding**

To create a set of time-domain optical fingerprints and build a library of different batches of $\tau^2$-Dots, we implemented five strategies to tailor the excited-state populations of emitters ($Tm^{3+}$ and $Er^{3+}$). As illustrated in Figure 2a, the strategies include the tuning of the core size, doping concentrations of emitters and sensitizer $Yb^{3+}$ in the core, the thickness of the core/sensitization layer, and the doping concentration of $Yb^{3+}$ in the sensitization layer, as well as the adding of a passivation layer. Using the five strategies (Table S1), fourteen (1→14 in Figure 2b), twelve (15→26 in Figure 2c), and sixteen (27→42 in Figure 2d) batches of three series of $\tau^2$-Dots have been synthesized in this work, which show finely tunable $\tau^2$-Dots profiles under NIR excitation at 976 nm or 808 nm. Though samples from the same doping series exhibit very similar emission colours, i.e., blue for the Yb-Tm series (Figure 2h), violetish blue for Nd-Yb-Tm series (Figure 2i), yellowish-green for the Nd-Yb-Er series (Figure 2j), their lifetime profiles display very differently in the time domain. Values in Figure 2e-2g further quantitatively map the large dynamic ranges of rising time, peak moment, and decay time distributions in identifying each batch of $\tau^2$-Dot samples.

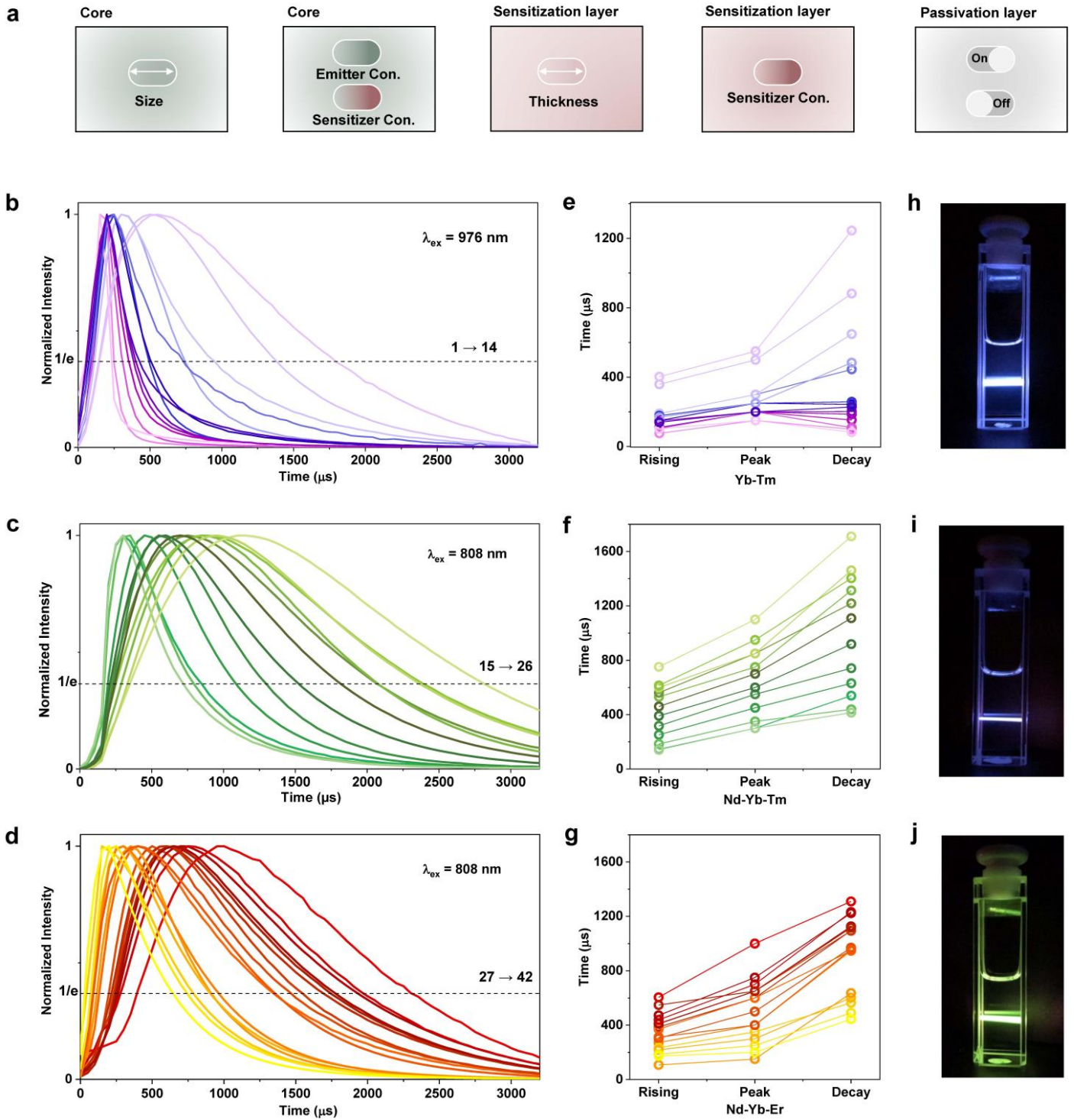

**Fig. 2 | Time-domain $\tau^2$-Dots profile control through upconversion energy transfer schemes and materials engineering. (a)** Illustrations of five strategies used for $\tau^2$-Dots profile tuning, i.e. core size, the concentrations of sensitizers and emitters in the core, the sensitization layer thickness, the concentration of sensitizers in the sensitization layer, and the passivation layer. **(b-d)** $\tau^2$-Dots profile tuning of three series of samples, i.e., Yb-Tm series (b), Nd-Yb-Tm series (c), and Nd-Yb-Er series (d), under NIR excitation. Dot lines indicate the normalized intensity of 1/e ($I_{1/e}$). **(e-g)** Calculated rising time ($\tau_{I_1} - \tau_{I_{1/e\_rising}}$), peak moment ($\tau_{I_1}$), and decay time ($\tau_{I_{1/e\_decay}} - \tau_{I_1}$) according to the curves in panels (b-d) for Yb-Tm series (e), Nd-Yb-Tm series (f), and Nd-Yb-Er series (g). **(h-j)** Photos of representative UCNPs in Yb-Tm series (h), Nd-Yb-Tm series (i), and Nd-Yb-Er series (j) showing their upconversion colours under NIR excitation.

## Optical uniformity of single $\tau^2$-Dots

Despite the large dynamic ranges of lifetime profiles can be encoded in different batches of $\tau^2$-Dots, the difference between each encoded optical fingerprints can be hidden at the ensemble level. Therefore, the single nanoparticle spectroscopy method has to be adopted to verify the optical uniformity of single $\tau^2$-Dot. Here, we selected fourteen batches of $\tau^2$-Dots (namely $\tau^2$-1 to $\tau^2$-14 in Table S2) in the Yb-Tm series ($\tau^2$-1 to $\tau^2$-9) and Nd-Yb-Er ($\tau^2$-10 to $\tau^2$-14) to perform the decoding experiment at single nanoparticle level. Using a confocal microscopy setup (Fig. S5), the single nanoparticle optical characterization result (Figure 3a and 3b) shows high degrees of brightness (e.g., 81,520 photon counts per second for $\tau^2$-13), optical uniformity (CV of 8.1%) (see other 4 batches of Nd-Yb-Er $\tau^2$-Dots in Fig. S6), and stability of single $\tau^2$-Dots (Figure 3c), ideal for long-term imaging and decoding of the optical fingerprint. As shown in Figure 3d, the unique and detectable fingerprint has been successfully assigned to every single $\tau^2$-Dot. More impressively, the characteristic lifetime fingerprints of single $\tau^2$-Dot, as long as from the same batch of synthesis, are consistently uniform.

**Wide-field time-resolved microscopy**

Confocal scanning microscopy allows illumination power up to $10^6$ W/cm$^2$ to excite every single nanoparticle by scanning across each pixel, but of which the scanning modality dramatically limits the throughput in the decoding process. We thus developed the wide-field microscope with an intensifier coupled CMOS camera for time-resolved imaging (Fig. S7). Under the wide-field microscopy, moderate continuous-wave excitation power density (5.46 kW/cm$^2$) sacrifices the brightness of each $\tau^2$-Dot by nearly two orders of magnitude (Fig. S8), but the wide-field microscopy enhances the decoding throughput by orders of magnitude, compared with the point scanning modality of confocal setup. As shown in Figure 3e, the sequence of time-resolved imaging consists of 75 frames (n=75), each recording the time-gated window period ($\Delta t$) of 50 μs.

**Nanoscale optical multiplexing of single $\tau^2$-Dots**

Compared to the conventional micron-sized beads, optical codes created on nanoscopic sized $\tau^2$-Dots can significantly increase the capacity of coding information, which takes optical super capacity multiplexing into the region smaller than the optical diffraction limit. To illustrate this opportunity and challenge, we stained 5 μm polystyrene beads with $\tau^2$-13 (Fig. S9) and collected their time-resolved upconversion images under a wide-field microscope. Within an illumination area of 28 μm in diameter, a typical image only contains less than ten micron-sized beads (Figure 3f), while in contrast, there are hundreds of single $\tau^2$-13 within the same area (Figure 3g). Each single micron bead shows a smooth $\tau^2$-Dot profile (Figure 3h), but the curve from a single $\tau^2$-Dot (Figure 3i) has some significant level of noise, due to the limited amount of detectable signal within each 50 μs time-gated window.

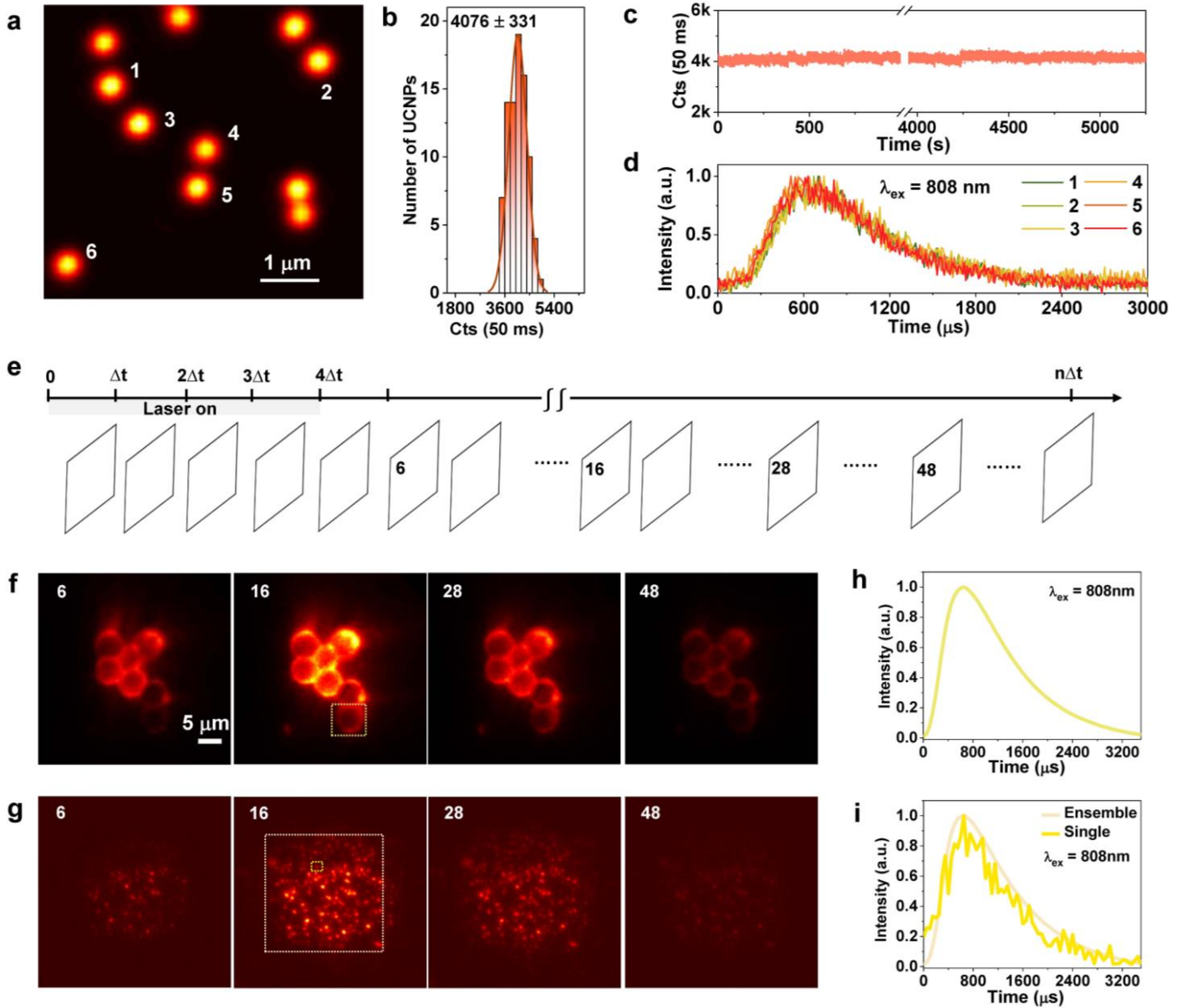

**Fig. 3 | Confocal and wide-field characterization of $\tau^2$-Dots. (a - d)** Confocal microscopic single nanoparticle imaging (a), brightness distribution (b), the long-term photostability of a single dot (c) under 808 nm CW excitation at $5.5 \times 10^6$ W/cm$^2$, and corresponding lifetime curves (d) of single dots 1-6 in (e) under 808 nm pulse excitation (by modulating the CW laser at 5.46 kW/cm$^2$). **(e)** Schematic illustration of the transient fluorescence signal detection principle using a time-resolved sCMOS camera for wide-field microscopy. **(f and g)** A comparison of the time-resolved 6$^{th}$, 16$^{th}$, 28$^{th}$, and 48$^{th}$ frames of $\tau^2$-Dots -stained micro-polystyrene beads (f) and single $\tau^2$-Dot (g) within a beam area of 28 μm in diameter. **(h)** Lifetime curve of a single $\tau^2$-Dot-stained bead, which is indicated by a dotted square in (f). **(i)** Lifetime curves of a single and ensemble of $\tau^2$-Dots monitored at 400-750 nm range, which are indicated by yellow and dark orange dotted squares in (g). All the data associate with a random batch of $\tau^2$-Dot ($\tau^2$-13).

## Extraction of high dimensional fingerprints

Using the wide-field time-resolved microscope, we detected the lifetime curves of single $\tau^2$-Dots from 14 batches. Though some detectable variations of the lifetime curves from dot to dot, caused by the illumination distribution (Fig. S10) and power-dependent intensities (Fig. S11), distinctive characteristics of each $\tau^2$-Dot and their lifetime tunability over a large dynamic range are clear (Figure 4a and 4d). Through the distribution statistics (Figure 4b and 4e), we find that most of the $\tau^2$-Dots have their $\tau_D$ (Decay indicator) values distributed uniformly with small CV (<10%, Fig. S12) and a small degree of overlap between each population, which is favourable for the decoding process. Four pairs of τ²-Dots populations, including τ²-2 vs τ²-3, τ²-4 vs τ²-5, τ²-8 vs τ²-9, and τ²-13 vs τ²-14, show significant overlappings. Strikingly, by adding one more indicator, extracted from the lifetime fingerprint profile, i.e., $\tau_R$ (Rising indicator), the two pairs of populations (τ²-4 vs τ²-5, τ²-8 vs τ²-9) could be well distinguished (Figure 4c). The remaining pairs remain too challenging to be separated even with the two indicators $\tau_D$ and $\tau_R$ (Figure 4c and 4f). These indicate that more high dimensional features are needed to be extracted from the lifetime profile for the individual nanoparticle identification.

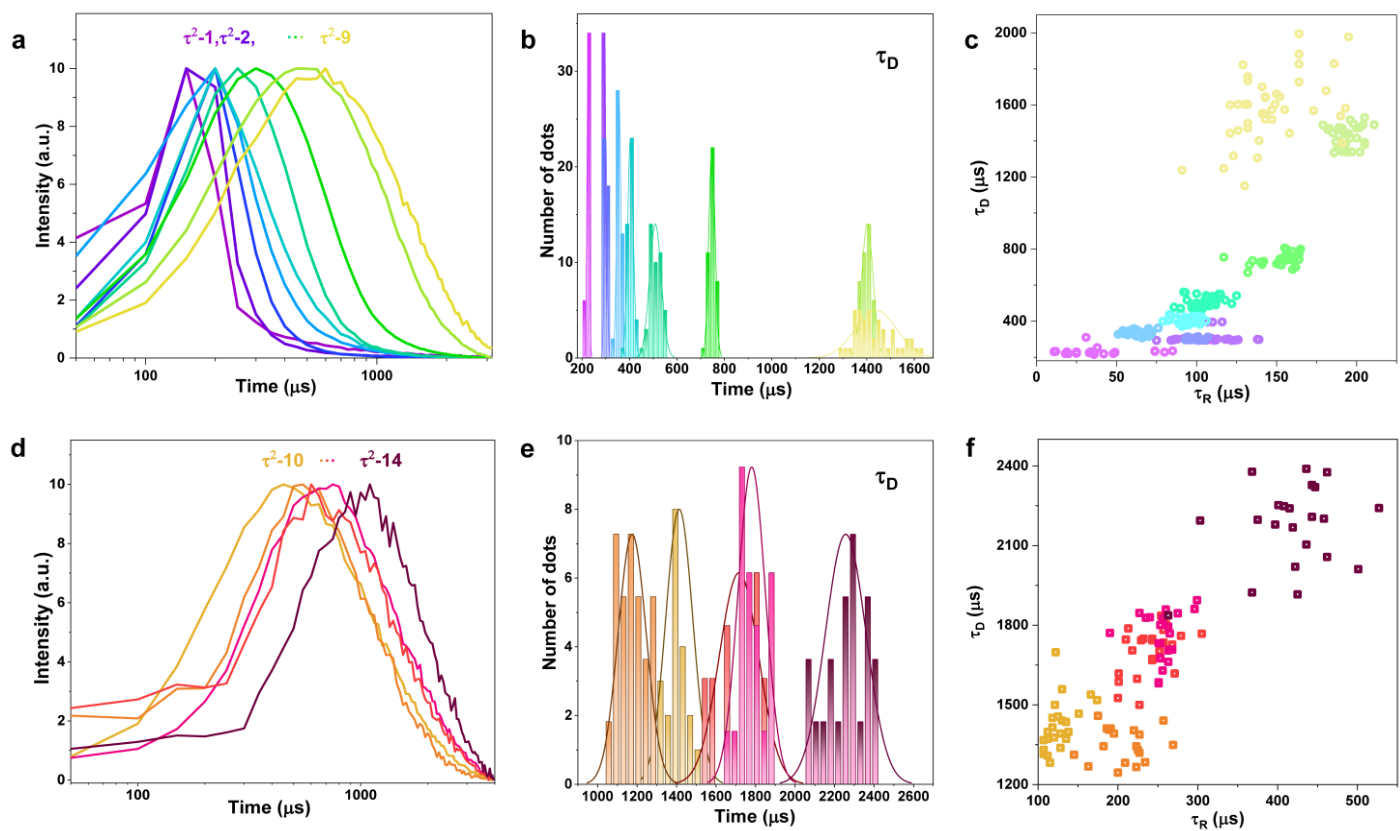

**Fig. 4 | Time-domain optical fingerprints from fourteen batches of τ²-Dots. (a and d)** Intensity normalized display of averaged single nanoparticle lifetime fingerprints of Yb-Tm series (nine) τ²-Dots and Nd-Yb-Er series (five) τ²-Dots. **(b and e)** The histograms of single-particle decay indicator ($\tau_D$) distribution analysis for the nine batches of Yb-Tm samples τ²-1 to τ²-9 (c) and the five batches of Nd-Yb-Er samples τ²-10 to τ²-14 (f). **(c and f)** Scatter plots of decay and rising indicators ($\tau_D$ and $\tau_R$) of samples τ²-1 to τ²-9 (d) and τ²-10 to τ²-14 (g). Both the indicators ($\tau_D$ and $\tau_R$) are defined as the time moment at 1/e of the maximum intensity.

**Deep learning approach**

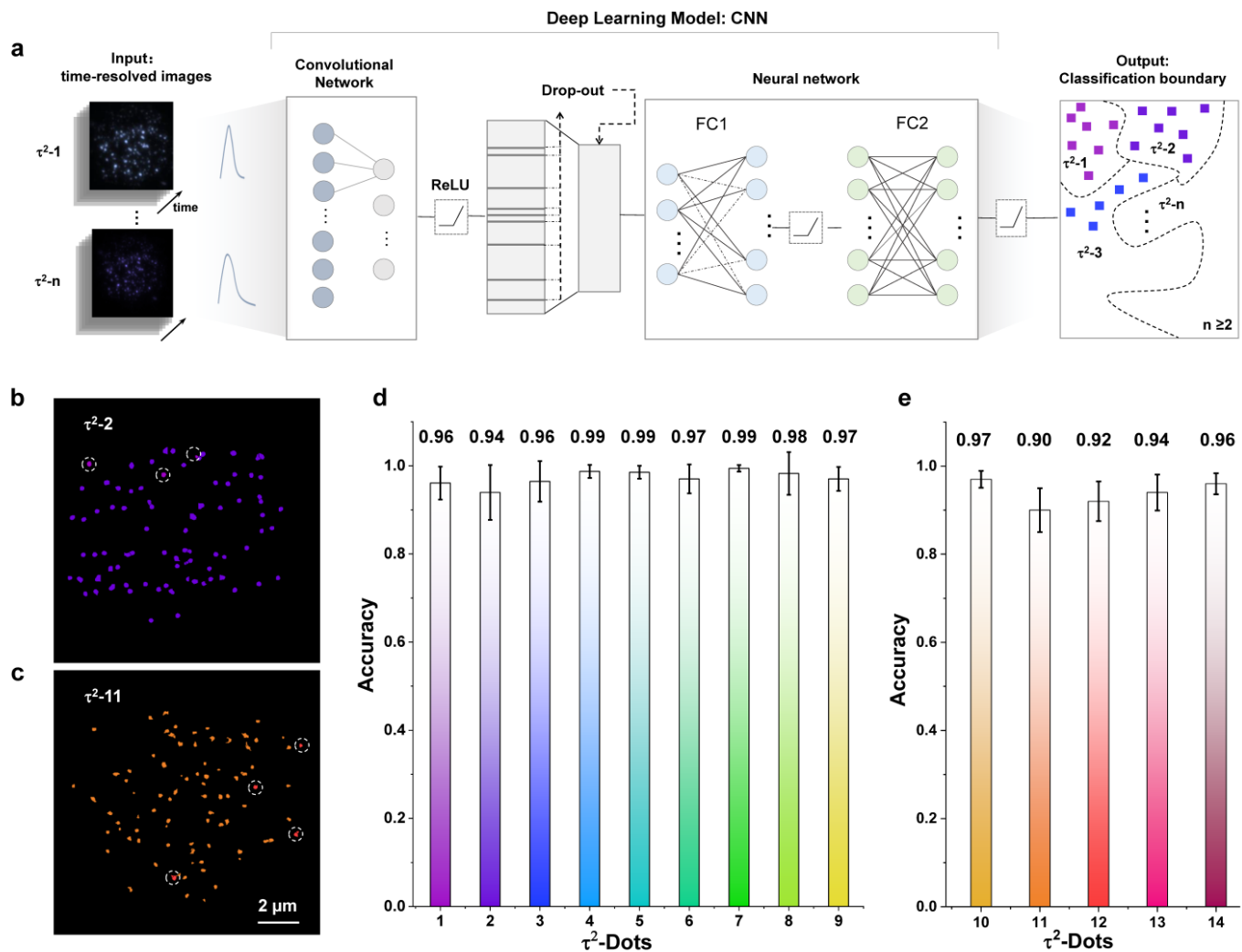

**Fig. 5 | Deep learning aided decoding of the fingerprints of single $\tau^2$-Dots. (a)** Illustration of the neural network used for the classification task. **(b and c)** One typical classification result for the $\tau^2$-2 (b) and $\tau^2$-11 (c) (for visualization purpose, pseudocolour is used to represent each type of single $\tau^2$-Dots). **(d and e)** Mean classification accuracy obtained through cross-validation with the database of 6 training sets and 1 validation set for each type of dots.

Deep learning is an emerging technique showing strong ability to classify highly non-linear datasets[50,51]. Here we show an opportunity offered by both the controlled growth of highly optically uniform single nanoparticles and subsequent image analysis to obtain their lifetime fingerprints of single dots, which can generate a large set of high-quality data to train the machine in deep learning. By collecting the sequences of time-dependent frames of images, we extracted the values of the normalized $\tau^2$-Dots profiles at 75 time moments between 0-3750 μs as the data source of input for training, in which we first pre-process the as-collected images by only selecting the imaging data from single nanoparticles (see more details in Methods, correlated SEM and optical image for single nanoparticle confirmation in Fig. S13, typical original images and single-particle images in Fig. S14). As shown in Figure 5a, we employed a convolutional network and a fully connected network with two layers (FC1 and FC2) to define the feature coverage for each batch of $\tau^2$-Dots (the classification boundaries).

We train the machine by the database of 14 batches of $\tau^2$-Dots with two series independently ($\tau^2$-1 to 9 and $\tau^2$-10 to 14) and challenge the established neural networks to recognize every single $\tau^2$-Dot. To do this, we first collected seven sets of time-resolved sequences of images from each type of $\tau^2$-Dot samples, and each image data contains the lifetime fingerprints of 50 to 200 single nanoparticles after data preselection of single $\tau^2$-Dots. We use any six sets of imaging data from each type of $\tau^2$-Dots to train the machine first to establish a neural network, and use the last set of data as validation analytes. A typical set of visualized results for $\tau^2$-2 and $\tau^2$-11 samples were displayed in Figure 5b and c . A small amount of mottled dots (e.g., in images of $\tau^2$-2 and $\tau^2$-11) represent the error recognition, which is mainly caused by the samples with similar lifetime curve features (Fig. S15). We then run the experiment of training and validation for another 50 times, each time randomly chose one set of data as the validation target and the other six sets to train the neural networks, which resulted in the statistical distributions of classification accuracy with error bars, displayed in Figure 5d and 5e. We achieved the mean classification accuracies for each $\tau^2$-Dot sample, with all the values approaching the unity. Table S3 and S4 show the high classification accuracy between two $\tau^2$-Dots samples over 20-trials under different neurons of per FC layers by 976 nm and 808 nm excitation, respectively. Their classification accuracies are hovering over the range of 91%-100%. The capacity of nanoscale multiplexing can be significantly determined by the brightness of single nanoparticles and the noise background, which explains the relatively broad distributions of $\tau^2$-Dot profiles for the batches of $\tau^2$-Dot samples with relatively low brightness, and therefore less accurate recognition results can be achieved by the machine intelligence (Fig. S16). Nevertheless, this experiment confirms the great potentials for the lifetime profiles of each $\tau^2$-Dot to be used for nanoscale super-capacity optical multiplexing, assisted by deep learning.

**Potential diverse applications using $\tau^2$-Dots**

The nanoscale super-capacity optical multiplexing opens a new horizon for many applications. Using the time-domain $\tau^2$-Dot profiles, different batches of materials emitting the same colour can be used to develop the new generation of dynamic anti-counterfeiting security inks, as illustrated in Figure 6a. Another unparalleled potential is to use nanoscale super-capacity multiplexing for high-throughput single molecular assay, which is superior to conventional suspension array assays based on microspheres. As a result of a proof of the principle experiment, in Figure 6b, we designed and functionalized (see Methods) the five kinds of $\tau^2$-Dots to simultaneously detect the five species of pathogenetic DNA sequences (Table S5) - hepatitis B virus (HBV), hepatitis C virus (HCV), human immunodeficiency virus (HIV), human papillomavirus type-16 (HPV-16), and Ebola virus (EV). Through a wide-field microscope, and compared with the control groups, we concluded that each $\tau^2$-Dots were highly specific. Moreover, as shown in Figure 5c, we demonstrate that the wide-field images of $\tau^2$-Dots with different lifetime profiles can be super-resolved using our latest development of upconversion structure-illumination microscopy (U-SIM) (Fig. S17)[52] with a resolution of 184.8 nm.

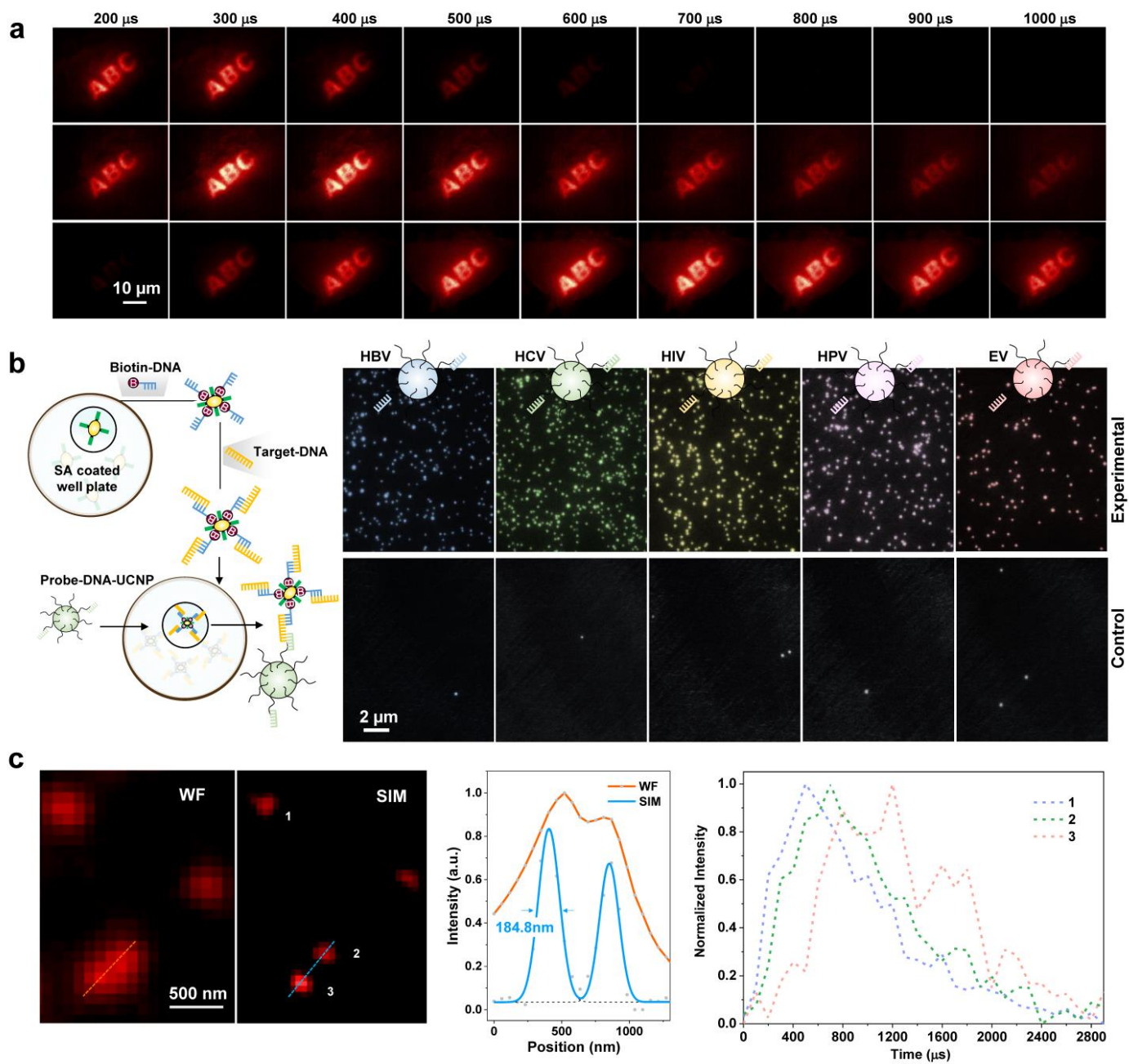

**Fig. 6 | Demonstration of the potentials of using the library of single $\tau^2$-Dots' optical fingerprints for a diverse range of applications. (a)** Time-domain anti-counterfeiting by using three types of Yb-Nd-Tm series of $\tau^2$-Dot security inks with different rising-decay fingerprints under 808 nm pulse laser excitation. **(b)** Multiplexed single molecule digital assays using five types of Yb-Tm series of $\tau^2$-Dot probes to quantify the five target pathogen single-strand DNAs (HBV, HCV, HIV, HPV-16, and EV) under 976 nm excitation. The cartoon illustration showing the probe-DNA conjugation procedure on a 96 well plate. **(c)** Three types Yb-Nd-Er series of $\tau^2$-Dots resolved by upconversion structure illumination microscopy (U-SIM).

## Conclusion

In this work, we realized the high throughput optical multiplexing at the nanoscale by coding and decoding the orthogonal fingerprints from single nanoparticles. This is based on using the sophisticated core-multi-shell design and controlled synthesis of bright and optically uniform upconversion nanoparticles. By controlling the energy transfer process, we have produced a total of 42 batches of nanoparticles and demonstrated the arbitrarily tuning of the time-domain $\tau^2$ profiles for each batch sample. We have further developed the wide-field imaging and deep-learning techniques to decode these sophisticated orthogonal fingerprints (excitation wavelength, emission colour and $\tau^2$-Dots profile) from single $\tau^2$-Dots in high throughput. Comparing to our previous work that uses the averaged decay time of UCNPs-decorated microspheres for multiplexing[44], the use of $\tau^2$-Dots profile from single nanoparticle and deep learning aided decoding process provides high-dimensional features for successful decoding at the nanoscale. This work also suggests that the concept for nanoscale multiplexing should also work for the other types of luminescent nanoparticles, as long as their optical uniformity and tunability of optical fingerprints, e.g. in the spectrum, meet the requirement discussed in this work. The encoding capacity will be enlarged by utilizing these orthogonal fingerprints and meet for future demands of high-throughput screening and super-resolution imaging of thousands of single molecule analytes in a single test.

## Methods

**Synthesis of UCNPs.** The NaYF$_4$ core nanoparticles were synthesized using a coprecipitation method[53]. In a typical procedure, 1 mmol RECl$_3$ (RE=Y, Yb, Nd, Er, Tm) with different doped ratios together with 6 mL oleic acid and 15 mL 1-octadecene were added to a 50 ml three-neck round-bottom flask under vigorous stirring. The resulting mixture was heated at 150 °C for 40 mins to form lanthanide oleate complexes. The solution was cooled down to 50 °C, and 6 mL methanol solution containing 2.5 mmol NaOH and 4 mmol NH$_4$F was added with vigorous stirring for 30 mins. Then the mixture was slowly heated to 150 °C and kept for 30 mins under argon flow to remove methanol and residual water. Next, the solution was quickly heated at 300 °C under argon flow for 1.5 h before cooling down to room temperature. The resulting core nanoparticles were collected and redispersed in cyclohexane with 5 mg/mL concentration after washing with cyclohexane/ethanol/methanol several times. We synthesized three series of core nanoparticles (NaYF$_4$: Yb, Tm; NaYF$_4$: Yb, Nd, Tm; NaYF$_4$: Yb, Er) with different doping concentrations using the same above method.

The shell precursors were prepared as the above procedure until the step where the reaction solution was slowly heated to 150 °C after adding NaOH/NH$_4$F solution and kept for 30 mins. Instead of further heating to 300 °C to trigger nanocrystal growth, the solution was cooled down to room temperature to yield the shell precursors.

The core-shell and core-multi-shell nanoparticles were prepared by layer by layer epitaxial growth method. The pre-synthesized NaYF$_4$ core nanoparticles were used as seeds for shell modification. 0.2 mmol as-prepared core nanocrystals were added to a 50 ml flask containing 3 ml OA and 8 ml ODE. The mixture was heated to 150 °C under argon for 10 min, and then further heated to 300 °C. Next, a certain amount of as-prepared shell precursors were injected into the reaction mixture and ripened at 300 °C for 2 mins, followed by the same injection and ripening cycles several times to get different shell thickness. Finally, the slurry was cooled down to room temperature and the formed core-shell nanocrystals were purified according to the same procedure used for the core nanocrystals. The core-multi-shell nanoparticles were also prepared by the epitaxial growth method described above and the core-shell nanoparticles were used as the seeds.

**Preparation of $\tau^2$-Dots ($\tau^2$-13) tagged microbeads.** The polystyrene (PS) microbeads (d=5 μm; Sigma-Aldrich) solution was processed by swelling 5 μl (10% W/v) of PS beads with 137 μl of an 8% (v/v) chloroform solution in butanol. 40 μl (8 mg/ml) $\tau^2$-13 dots in cyclohexane was added to the above PS suspension. The solution was vortexed after adding the $\tau^2$-Dots. After incubating at 25 °C for 3 hours, the beads were washed four times, alternating between ethanol and cyclohexene. After washing, the $\tau^2$-Dots embedded beads were dispersed in ethanol and then one drop of the beads was air-dried on the surface of a coverslip for optical measurements.

**Material characterizations.** The morphology characterization of the nanoparticles was performed by transmission electron microscopes of JEOL TEM-1400 at an acceleration voltage of 120 kV and JEOL TEM-2200FS with the 200 kV voltage. The high-resolution TEM (HTEM) images were performed on the Helios G4 PFIB series UXe Dual Beam systems with an accelerating voltage of 30 kV. The cyclohexane dispersed UCNPs were imaged by dropping them onto carbon-coated copper grids. The surface morphology characterization of the PS beads (Fig. S9) and the light-electron microscopic correlation experiment (Fig. S13) were performed by using a Zeiss Supra 55VP Scanning Electron Microscope (SEM) operated at 20.00 kV.

**Preparation of sample slides.** To prepare a sample slide for single nanoparticle measurement, a coverslip was washed with pure ethanol by ultrasonication, followed by air-drying. 10 μl of the $\tau^2$-Dots (0.01 mg/ml) in cyclohexane was dropped onto the surface of a coverslip. After being air-dried, the coverslip was put over a clean glass slide and any air bubbles were squeezed out by gentle force before measurement.

**Confocal imaging and lifetime measurement.** We built a stage-scan confocal microscope for the intensity and lifetime measurements of single $\tau^2$-Dots, as shown in Fig. S5. The excitation source of 808 nm single-mode polarized laser was focused onto the sample through a 100x objective lens (UPLanSApo100X, oil immersion, NA = 1.40, Olympus Inc., JPN). The emission from the $\tau^2$-Dots sample was collected by the same objective lens and refocused into an optical fibre which has a core size matching with the first Airy disk of the system. The fluorescence signals were filtered from the laser by a short-pass dichroic mirror (DM, ZT785spxxr-UF1, Chroma Inc., USA) and a short pass filter (SPF, ET750sp-2p8, Chroma Inc., USA). A single-photon counting avalanche photodiode (APD, SPCM-AQR-14-FC, Excelitas Inc., USA) was connected to the multi-mode fibre (MMF, M42L02, Thorlabs Inc., USA) to detect the emission intensity. The scanning was achieved by moving the 3D piezo stage. Every single nanoparticle showed a Gaussian spot in the confocal scanning microscopic image. The maximum brightness value (photon counts) of each Gaussian spot was used to represent the brightness of that single nanoparticle. We evaluated more than 20 single nanoparticles to calculate the mean brightness.

For the lifetime measurement, we modulated the diode laser to produce 200 μs excitation pulses. The photon-counting SPAD was continuously switched on to capture the long-lifetime luminescence. For each time point, the gate-width is 50 μs with an accumulation of 10000 times. The pulsed excitation, time-gated data collection, and the confocal scanning were controlled and synchronized using a multifunction data acquisition device (USB-6343, National Instruments) and a purpose-built LabVIEW program.

**Wide-field spectrum and lifetime measurement.** We built a wide-field fluorescence microscope, as shown in Fig. S7, to acquire the fluorescence lifetime image sequences of $\tau^2$-Dots. A single-mode diode-pumped solid-state laser (LU0808M250, Lumics Inc., GER, 808 nm, the excitation power density of 5.46 kW/cm$^2$) was used to excite the $\tau^2$-Dots after expanding the laser beam by three times. The emission of $\tau^2$-Dots was collected by a high NA objective lens (UPLanSApo100X, oil immersion, NA = 1.40, Olympus Inc., JPN) and separated from the laser reflection by a short-pass dichroic mirror (DM1, ZT785spxxr-UF1, Chroma Inc., USA) and a short pass filter (SPF, ET750sp-2p8, Chroma Inc., USA), then focused by a tube lens to the time-resolved sCMOS camera (iStar sCMOS, Andor Inc., UK). The camera also functions as the pulse modulator of an exciting laser beam via a BNC cable. By applying the Kinetics Mode of the camera and Integrate-On-Chip (IOC) at 250Hz, we acquired the lifetime image sequences of 75 frames from 0 μs to 3750 μs with a time gate of 50 μs, under the laser excitation pulse of 0-200 μs. The IOC mode enabled the accumulation of fluorescence signal with the greatly improved signal-to-noise ratio. To measure the fluorescence lifetime image sequences of $\tau^2$-Dots under 976 nm excitation, a single-mode 976 nm laser (BL976-PAG900, Thorlabs Inc., USA, the excitation power density of 8.7 kW/cm$^2$) was added in the setup as the excitation light. After collimation, the excitation beam was expanded by 2.5 times and then reflected by the short-pass dichroic mirror (DM2, T875spxrxt-UF1, Chroma Inc., USA), and focused through the objective lens to the sample slide. The fluorescence signals can also be coupled into a multi-mode fibre (MMF, M24L02, Thorlabs Inc., USA) by switching a flip mirror and then detected by a miniature monochromator (iHR550, Horiba Inc., JPN) for measuring upconversion emission spectra. The spectral region ranged from 400 to 750 nm.

**Data processing and networks for deep learning.** To perform single nanoparticle-based machine learning, we first performed data processing to mark the single nanoparticles in the collected images. We first selected the brightest frame (maximum mean brightness) from the 75 frame images. Then we found the peak pixel of each bright spot. For each peak, we cropped a 40-pixel by 40-pixel region of interest (ROI) centred on the peak. In each ROI, we segmented the image with the OTSU threshold and got a binary mask. Considering that two adjacent peaks might be connected in the binary mask, we employed watershed segmentation on the binary mask to get the boundaries of each peak. Finally, we sorted all the spots by its peak intensity and divide all spots into four groups (Q1 to Q4) according to their peak intensities. The Q1-Q4 represented 4 intensity thresholds to classify the groups. We counted the spots as the single nanoparticles when the peak intensities within the statistical range of single particle intensity (eg. 8000±1000 for $\tau^2$-13, equaling to Q2 group). After filtering out all the aggregated spots, we obtained the image that only involves single nanoparticles. After that, the image sequence was transformed into multiple single nanoparticle sequences. For example, if 100 particles were identified as single nanoparticles in an image sequence, this image sequence was decomposed into 100 particle sequences.

The artificial neural networks (ANN) were implemented in python using the PyTorch package. (https://pytorch.org/). We extracted the normalized time-domain fluorescence intensity sequences of single nanoparticles as the input for deep learning. We performed the aforementioned data processing for all the 14 $\tau^2$-Dots. We randomly picked out ~500 single nanoparticles for each $\tau^2$-Dots as the training sets, where their lifetime features and types were known. ~ 100 single nanoparticles were used as the validation sets. There were five key aspects while determining the networking architecture: 1) the number of layers in the convolutional network; 2) the number of filters in each 1D convolutional layer; 3) whether to use activation function; 4) the number of neurons in each fully connected (FC) layer; 5) the keep probability for the dropout regularization scheme. We started with the network structure of one convolutional layer with 10 filters and two fully connected layers with 10 neurons for each of them. Fig. S18 shows the categorical cross-entropy loss as increasing training epochs, which shows that all the parameters were well-optimized.

We first determined the number of neurons in each fully connected layer ranging from 10 to 1000. Given one convolutional layer with 10 filters, the network obtained satisfactory results when the number of neurons in each FC layer was around 500. Given the above two FC layers, we started to determine the number of convolutional layers and the number of filters for each layer. The network obtained satisfactory results when using two convolutional layers with 30 filters in the first layer and 20 filters in the second layer. Then, given the above convolutional layers, we further adjusted the number of neurons for each FC layer, and found 10-80 neurons in each layer can obtain satisfactory results. With the above conductions, the network structure was temporarily determined as two convolutional layers with 30 filters in the first layer and 20 filters in the second layer followed by two FC layers with 50 and 30 neurons for each layer. The plotting validation curves for the number of neurons for FC1 and FC2 layer by training and testing 20 times were shown in Fig. S19 and S20, in which each training-testing randomly selects samples from the single particle database. With this network structure, we validated the network performance when activation functions or/and dropout scheme was/were introduced. Three activation functions have been validated during this procedure, which were ReLU, ReLU6 and RReLU. The validation curve of the dropout rate of the network structure was shown in Fig. S21, the keep probability of the dropout scheme was determined in the range from 0.1 to 0.9. After the above adjustment of the network, we went back to adjust the number of neurons in FC layers and obtained the final network architecture as below.

The fingerprint retrieval network contained two convolutional networks and two fully connected networks. The two 1D convolutional layers used the element-wise function $ReLU6(x) = \min(\max(0,x),6)$. There were 50 filters in the first 1D convolutional layer using a kernel of size 2 and the stride size was 2. The second 1D convolutional layer has 20 filters with a kernel of size 2 and the stride size was 1. The two fully connected networks contained two layers with 50 (FC1) neurons in the first layer and 30 (FC1) neurons in the second layer, and the element-wise function was also employed for each layer. We applied a dropout regularization scheme with 80% keep probability for the fully connected part (i.e. the dropout rate is 0.2). During training, the output layer neuron whose index corresponds to the input binary number was set to "1" while the other neuron activations were kept at "0". A variant of the stochastic gradient descent (SGD) algorithm ("Adam") was applied to train the parameters in the network through a randomly shuffled batch size 256 (Fig. S22). We used a learning rate of 0.005 and trained the network for 50 epochs. The validation curves of the learning rate was shown in Fig. S23.

The classification effectiveness of convolutional neural networks was evaluated by the mean and deviation of the classification accuracy of 50 randomly sampled experiments. We have 7 sets of image sequences of each sample and run 50 experiments of training-and-testing to compute the average error and deviation. In each experiment, we randomly selected one set of image sequences for test particles. For 14 batches of nanoparticles, we selected 14 image sequences. The data of single nanoparticles in the rest image sequences were used as the training set in the training algorithm section, where their lifetime features were available, but the label was unknown until computing the model error. After one training-and-testing process, the testing error for 14 image sequences was obtained. The mean and deviation of errors were computed through 50 random selections.

**Anti-counterfeiting experiment.** The time-domain anti-counterfeiting by using three types of Yb-Nd-Tm series of $\tau^2$-Dots was based on the spatial modulation of the excitation patterns on the sample plane. A digital micro-mirror device (DMD) was added in the wide-field optical system as the spatial light modulator to generate excitation patterns of the ABC alphabet. The laser beam illuminated the DMD after beam collimation and expansion. Then the illuminated alphabet patterns were imaged on the sample plane. Then we acquired the lifetime image sequences of these three batches of $\tau^2$-Dots under 808 nm pulse laser illumination of ABC alphabetical patterns.

**DNA assay experiment.** Post-synthesis surface modification was adopted to transfer the $\tau^2$-Dots into hydrophilic and biocompatible before bioconjugation with DNA oligonucleotides. Surface modification was performed *via* ligand exchange with block copolymer composed of hydrophilic block poly(ethylene glycol) methyl ether acrylate phosphate methacrylate (POEGMA-b-PMAEP)[54]. In a typical procedure, 500 μL of OA-coated $\tau^2$-Dots (20 mg/mL) were dispersed in tetrahydrofuran (THF). Then the OA-capped $\tau^2$-Dots in THF were mixed with 5 mg copolymer ending in carboxyl group in 2 mL THF. The above mixture was sonicated for one min followed by incubation in a shaker overnight at room temperature. The polymer-coated $\tau^2$-Dots were purified four times by washing/centrifugation at 14,860 rpm for 20 min with water to obtain carboxyl group modified $\tau^2$-Dots. The supernatant was removed and the nanoparticles were dispersed in water for further conjugation with DNA sequences.

We selected five couples of pathogen-related genetic sequences in the short length of 24 bases (HBV, HCV, HIV, HPV-16, EV), the sequences of these oligonucleotides used in this work were listed in Table S5. The protocol of carbodiimide chemistry was adopted to conjugate the carboxyl group on the polymer with the amine groups of probe DNA molecules. The five groups of carboxyl-$\tau^2$-Dots were pre-activated by the EDC (100-folder molar ratio to carboxyl-$\tau^2$-Dots) in HEPES buffer (20 mM, pH 7.2) with slightly shaking at room temperature for 30 mins. The five groups of NH$_2$-DNA (100 μM) was added into the above solution with 600 rpm shaking for the reaction of 3 h, respectively. The activated carboxyl-$\tau^2$-Dots were washed/centrifuged at 14680 rpm cycle two times to remove EDC and resuspended in HEPES buffer to obtain probe DNA-polymer-$\tau^2$-Dots.

The streptavidin with a concentration of 10 μg/mL in 100 μL PBS buffer was coated on the 96-well plates and incubated 4 h at room temperature. Following by removal of the supernatant, 200 pmol biotinylated-capture DNA in 200 μL PBS was added into the well and incubated overnight at 4°C for further immobilization. Washing the plates three times with PBS buffer after the reaction, then 200 μL of blocking with 1% casein buffer was added to each well and incubated at room temperature for 1 h. The Target-DNA in 200 μL Tris buffer contains 0.1% casein was added to the five experimental wells individually and incubated at room temperature for 2 h, while the five corresponding control wells were added with the same reaction buffer without Target-DNA. After washing three times with Tris buffer, 100 μL complementary DNA-functionalized $\tau^2$-Dots in reaction buffer contains 0.1% casein and 5 mM NaF in Tris were added to react 1 h. Then washing the wells three times and the well was ultimately dissolved in 100 μL Tris-5mM NaF before detecting the images.

**SIM imaging experiment.** Structured illumination microscopy (SIM), as a wide-field super-resolution technique, was based on the spatial modulation of the excitation patterns on the sample plane. In this work, a digital micro-mirror device (DMD, DLP 4100, Texas Instruments Inc., USA) was used as the spatial light modulator to generate excitation patterns. DMD contained an array of 1024×768 micro-mirrors on the chip. The size of each micromirror was 13.68×13.68 μm$^2$. Each micro-mirror can be tilted to two positions along its diagonal: ±12° tilt to deflect the incident light beam away from the optical path. These micro-mirrors can be controlled independently to modulate the amplitude of incoming light to generate arbitrary illumination patterns.

As shown in Fig. S17, the optical system for the time-resolved SIM was built based on conventional widefield fluorescence microscopy (Fig. S7) with proper modification. Then we acquired the nine groups of lifetime image series with nine illuminating patterns, corresponding to three different angular orientations ($\theta1 = 0°$, $\theta2 = 60°$ and $\theta3 = 120°$) and three different phase shifts ($\varphi1 = 0°$, $\varphi2 = 120°$ and $\varphi3 = 240°$). The other test parameters were the same as the conventional widefield microscopy. In the reconstruction of the super-resolution image series, these nine images are processed in the Fourier domain to reconstruct the final super-resolution image. As shown in Figure 6c, three batches of $\tau^2$-Dots ($\tau^2$-10, $\tau^2$-12, and $\tau^2$-14) were mixed for imaging. Single dots 2 and 3 owning different lifetime curves were separated under the SIM modality by comparing the brightest frames from time-resolved conventional widefield microscopy and SIM.

**Data Availability**

The datasets including npy and mat formats of 14 kinds of $\tau^2$-Dots used for deep learning are available at Supplementary Dataset.

**Code Availability**

The pre-trained neural network code used in this article is available at Supplementary Dataset.

**Acknowledgments:** The authors acknowledge the financial support from the Australian Research Council Discovery Early Career Researcher Award Scheme (J. Z., DE180100669), China Scholarship Council Scholarships (Jiayan Liao: No. 201508530231, Baolei Liu: No.201706020170, Chaohao Chen：No. 201607950009), Australia-China Joint Research Centre for Point-of-Care Testing (ACSRF65827, SQ2017YFGH001190), Science and Technology Innovation Commission of Shenzhen (KQTD20170810110913065), National Natural Science Foundation of China (61729501), and Major International (Regional) Joint Research Project of NSFC (51720105015).

**Author contributions:** J.Z., and D.J. conceived the project, designed the experiments and co-supervised the research; J.Liao conducted the synthesis, characterization and measurement. J.Liao and B.L. processed the data; J.Liao and Y.C performed the single molecule assay. Y.S. and J. Lu conducted machine learning experiment; B.L., F.W., J.Liao and C.C. built the optical system; J.Liao and J.Z. prepared the figures and Supplementary Materials; J.Z. and D.J. wrote the manuscript with input from other authors.

**Competing interests:** The authors declare no competing financial interest.